\documentclass[aps,prd,floats,floatfix,nofootinbib,superscriptaddress]{revtex4-1}
\usepackage{graphicx,amsmath}
\usepackage{amssymb}
\usepackage{amsfonts}

\begin{document}

\title{Model-independent Higgs coupling measurements at the LHC \\ 
using the $H \to ZZ \to 4 \ell$ lineshape}

\author{Heather E.~Logan}
\email{logan@physics.carleton.ca}
\affiliation{Carleton University, Ottawa, Ontario K1S 5B6, Canada}

\author{Jeff Z.~Salvail}
\email{jsalv039@uottawa.ca}
\affiliation{University of Ottawa, Ottawa, Ontario K1N 6N5, Canada}

\date{July 21, 2011}

\begin{abstract}
We show that combining a direct measurement of the Higgs total 
width from the $H \to ZZ \to 4 \ell$ lineshape with Higgs signal rate measurements allows Higgs couplings
to be extracted in a model-independent way from CERN Large Hadron Collider 
(LHC) data.  Using existing experimental studies with 30~fb$^{-1}$ at one detector of 
the 14~TeV LHC, we show that the couplings-squared of a 190~GeV Higgs
to $WW$, $ZZ$, and $gg$ can be extracted with statistical precisions of about 
10\%, and a 95\% confidence level upper limit on an unobserved component 
of the Higgs decay width of about 22\% of the SM Higgs width can be set.
The method can also be applied for heavier Higgs masses.
\end{abstract}

\maketitle

\section{Introduction}

The Higgs mechanism for electroweak symmetry breaking and generation of fermion masses remains the last untested component of the Standard Model (SM).  The discovery of the physical Higgs boson that accompanies this mechanism, or its refutation through the discovery of alternative dynamics underlying electroweak symmetry breaking, is the primary goal of the CERN Large Hadron Collider (LHC).
The SM Higgs mechanism predicts the couplings of the Higgs boson to pairs of SM particles in terms of the known particle masses.  In extensions of the SM Higgs sector, these couplings
are typically modified, with the pattern of modifications providing valuable information that can
shed light on the structure of the extended model.  A key test of the SM Higgs mechanism
thus involves measurement of as many Higgs couplings as possible.

High-precision, model-independent measurements of Higgs couplings are a major component
of the physics case for the International Linear $e^+e^-$ Collider (ILC)~\cite{Djouadi:2007ik}.  Key to the model-independence of these measurements is the ability to measure the total production cross section for $e^+e^- \to Z H$ using the recoil-mass technique~\cite{GarciaAbia:1999kv}.  This technique requires event-by-event knowledge of the four-momentum of the initial state (without using measurements of the Higgs decay products), and is thus unavailable at the LHC.  Instead, extraction of Higgs couplings from future LHC measurements of Higgs production and decay rates requires making a model-dependent assumption, either about the possible Higgs decay modes or about some of the Higgs couplings.

The difficulty arises due to a genuine flat direction in the coupling fit corresponding to allowing an unobserved Higgs decay mode while simultaneously increasing all the production and decay couplings by a common factor.  The Higgs signal rate in production channel $i$ and decay channel $j$ is given by
\begin{equation}
	{\rm Rate}_{ij} = \sigma_i \, {\rm BR}_j = \sigma_i \frac{\Gamma_j}{\Gamma_{\rm tot}},
	\label{eq:rate1}
\end{equation}
where $\sigma_i$ is the cross section for production channel $i$, BR$_j$ is the Higgs branching ratio into final state $j$, $\Gamma_j$ is the Higgs partial width into final state $j$, and $\Gamma_{\rm tot}$ is the Higgs total width.  Adding a new, unobserved Higgs decay channel with partial width $\Gamma_{\rm new}$ while simultaneously increasing all the Higgs couplings by a common factor $a$ relative to their SM values yields 
\begin{equation}
	{\rm Rate}_{ij} = a^2 \sigma_i^{\rm SM} 
	\frac{a^2 \Gamma_j^{\rm SM}}{a^2 \Gamma_{\rm tot}^{\rm SM} + \Gamma_{\rm new}}.
	\label{eq:rate2}
\end{equation}
For any given value of $\Gamma_{\rm new}$, an appropriate choice of $a$ yields Higgs signal rates identical to those in the SM.

Previous studies of Higgs coupling extraction from LHC data have dealt with this flat direction either by assuming that no unexpected Higgs decay channels exist~\cite{Zeppenfeld:2000td,Belyaev:2002ua,Lafaye:2009vr}, or by assuming that the Higgs couplings to $WW$ and $ZZ$ are bounded from above by their SM value~\cite{Duhrssen:2004cv}.\footnote{The latter assumption holds in any extended Higgs sector that contains only regular SU(2)$_L$ doublets and singlets.  Viable models in which it is violated include the Georgi-Machacek model with SU(2)$_L$-triplet Higgses~\cite{Georgi:1985nv,Chanowitz:1985ug} and the Lee-Wick Standard Model~\cite{Grinstein:2007mp} which solves the hierarchy problem by implementing Pauli-Villars regularization with physical fields~\cite{Lee:1969fy}.  The potential enhancements of the Higgs couplings to $WW$ and $ZZ$ above their SM values in these models have been studied in Refs.~\cite{Logan:2010en} and \cite{Alvarez:2011ah}, respectively.}  These analyses have focused on Higgs masses between 100 and 190~GeV where many different Higgs production and decay modes are experimentally accessible.  Depending on the Higgs mass, they have shown that LHC measurements will provide sensitivity to the Higgs couplings to $W$ and $Z$ pairs, top and bottom quarks, and tau leptons, as well as to potential new loop contributions to the Higgs couplings to photon and gluon pairs.  Theoretical and systematic uncertainties play an important role in the fits~\cite{Duhrssen:2004cv,Lafaye:2009vr}, and in general the extracted coupling values are correlated with one another~\cite{Lafaye:2009vr}.

In this paper we propose a method by which the model dependence can be removed from Higgs coupling measurements at the LHC.  When the total width of the Higgs is above about a GeV, it can be measured directly using the lineshape of the four-lepton invariant mass distribution in $H \to ZZ \to 4 \ell$, where $\ell = e$,~$\mu$.  This occurs in the SM for Higgs masses above about 190~GeV.\footnote{The global fit to precision electroweak observables in the SM puts an upper bound on the SM Higgs mass of 169~(200)~GeV at 95\%~(99\%)~confidence level~\cite{Baak:2011ze}.  Thus, Higgs masses heavy enough to resolve the Higgs total width from the $H \to ZZ \to 4\ell$ lineshape are disfavored in the context of the SM.  However, in extended models, Higgs masses above 190~GeV are fully consistent with electroweak precision measurements; see Ref.~\cite{Baak:2011ze} for an extensive review.  Therefore, if such a heavy Higgs boson is discovered, the electroweak fit provides strong motivation to search for new physics effects, including shifts in the Higgs couplings relative to the SM expectations.}  Such a measurement directly determines the denominator in Eqs.~(\ref{eq:rate1}) and (\ref{eq:rate2}), removing the degeneracy in the fit without imposing any model assumptions.  

The rest of this paper is organized as follows.  In Sec.~\ref{sec:param} we describe our parameterization of possible new physics in the Higgs couplings.  In Sec.~\ref{sec:observables} we list the Higgs observables that we use in the fit and describe the fitting procedure.  For a Higgs mass of 190~GeV, current LHC studies indicate that the only observable SM Higgs production modes will be gluon fusion and vector boson fusion and the only observable SM decays will be to $WW$ and $ZZ$.  Our fit is thus sensitive only to the Higgs couplings to $WW$ and $ZZ$, to the effective Higgs coupling to two gluons, and to a possible new component of the Higgs total width.  In Sec.~\ref{sec:results} we present the results of our fits as chi-squared contours in two-dimensional projections of our parameter space, as well as giving 1- and 2-sigma constraints on the couplings, assuming the best-fit point is at their SM values.  In Sec.~\ref{sec:conclusions} we discuss our results and conclude.

\section{Parameterization of Higgs couplings}
\label{sec:param}

A SM Higgs boson with mass 190~GeV decays 78.70\% of the time to $W^+W^-$ and 20.77\% to $ZZ$, with the remaining 0.53\% of decays mainly to $b \bar b$ and $gg$, leading to a total width of  1.036~GeV (computed using the public FORTRAN code {\tt HDECAY}~\cite{HDECAY}).  The largest production cross sections for such a Higgs boson at the LHC are gluon fusion ($gg \to H$) and vector boson fusion (VBF $\to H$).  At this mass, LHC measurements will provide access to the Higgs total width through the $H \to ZZ \to 4\ell$ lineshape, as well as event rates in four primary Higgs production and decay channels: $gg \to H \to WW$, $gg \to H \to ZZ$, ${\rm VBF} \to H \to WW$, and ${\rm VBF} \to H \to ZZ$.  LHC measurements can thus provide sensitivity to the Higgs couplings to $WW$ and $ZZ$ (through Higgs decays to these final states and through production via vector boson fusion), the effective coupling to $gg$ (through production via gluon fusion), and potential new contributions to the Higgs total width (directly through the $H \to ZZ \to 4 \ell$ lineshape as well as indirectly through the Higgs branching ratios to $WW$ and $ZZ$ final states).  

We parameterize shifts in these three couplings in terms of three multiplicative factors, $\bar g_W$, $\bar g_Z$, and $\bar g_g$, which are all equal to $1$ in the SM.  We parameterize shifts in the Higgs total width in terms of a new, unobserved contribution to the width, $\Gamma_{\rm new}$, which is zero in the SM.  In terms of these parameters, the Higgs partial and total widths and production cross sections are given as follows (we neglect SM contributions to $\Gamma_{\rm tot}$ other than $WW$ and $ZZ$):
\begin{eqnarray}
	\Gamma_W &=& \bar g_W^2 \Gamma_W^{\rm SM}, \nonumber \\
	\Gamma_Z &=& \bar g_Z^2 \Gamma_Z^{\rm SM}, \nonumber \\
	\Gamma_{\rm tot} &=& \Gamma_W + \Gamma_Z + \Gamma_{\rm new}, \nonumber \\
	\sigma(gg \to H) &=& \bar g_g^2 \sigma^{\rm SM}(gg \to H), \nonumber \\
	\sigma({\rm VBF} \to H) &\simeq& \bar g_W^2 \sigma_W^{\rm SM} + \bar g_Z^2 \sigma_Z^{\rm SM} \simeq \left[ 0.73 \, \bar g_W^2 + (1 - 0.73) \bar g_Z^2 \right] \sigma^{\rm SM}({\rm VBF} \to H).
	\label{eq:params}
\end{eqnarray}

Two comments are in order.  First, vector boson fusion proceeds through diagrams involving either $W$ or $Z$ exchange in the $t$-channel.  If we take $\bar g_W = \bar g_Z \equiv \bar g_V$ as predicted in most extended Higgs models, then the cross section for vector boson fusion is given simply by $\sigma({\rm VBF} \to H) = \bar g_V^2 \sigma^{\rm SM}({\rm VBF} \to H)$.  If, however, we want to fit separately for $\bar g_W^2$ and $\bar g_Z^2$ (which are unequal, for example, in the Georgi-Machacek model~\cite{Georgi:1985nv,Chanowitz:1985ug} with custodial SU(2) violation in the Higgs mixing~\cite{Logan:2010en}), we must express the vector boson fusion cross section in terms of the separate contributions of the $W$ and $Z$ exchange diagrams.  These diagrams do interfere, e.g., in $ud \to udH$; however, the interference is negligible as can be understood by examining the kinematics of the $W$ and $Z$ exchange diagrams: in $W$ exchange the final-state $u$ quark is produced primarily in the same direction as the initial $d$ quark, while in $Z$ exchange it is produced primarily in the opposite direction.
We have checked this numerically using the public code {\tt MadGraph}~\cite{MadGraph} by computing the tree-level cross section for $pp \to jjH$ and comparing it to the sum of the cross section including only $W$ exchange and the cross section including only $Z$ exchange.  From this calculation we obtain the factors of $0.73$ and $(1 - 0.73)$ for the $W$ and $Z$ fractions of the vector boson fusion cross section quoted in the last line of Eq.~(\ref{eq:params}).

Second, the LHC measurements that we consider do not provide access to Higgs couplings to fermions.  The cross section for $gg \to H$ is sensitive to the Higgs coupling to the top quark, which dominates the one-loop gluon fusion process in the SM; however, new colored particles running in the loop can also affect $gg \to H$ and their effects cannot be disentangled from a shift in the Higgs coupling to the top quark based on this single measurement.  
A large enhancement of the Higgs coupling to bottom quarks can lead to a non-negligible branching fraction for $H \to b \bar b$; we capture this possibility here only in $\Gamma_{\rm new}$.  A large enhancement of the Higgs coupling to bottom quarks can also affect the $gg \to H$ cross section through the contribution of the bottom quark in the loop.  In this case, tree-level Higgs production through $b \bar b \to H$ can also become a significant contributor to the inclusive Higgs production cross section, so that the value of $\bar g_g^2$ extracted through our fit will be contaminated with this additional production mode.  Searching for $gb \to Hb$ production with an extra tagged $b$ quark in the final state could help pin down this scenario.  We do not pursue this possibility further here.

Starting from the SM prediction for the event rate in each channel and the Higgs total width, the parameterization in Eq.~(\ref{eq:params}) lets us recompute these rates and total width for any point in the four-dimensional parameter space of $\bar g_W^2$, $\bar g_Z^2$, $\bar g_g^2$, and $\Gamma_{\rm new}$.  By comparing these model predictions to the expected experimental precision on the observables, we will determine the precision with which the four parameters can be measured.

\section{Observables and fitting procedure}
\label{sec:observables}

We now describe the observables used in our fit and the fitting procedure.  Our fit involves seven observables, comprising the Higgs total width and event rates in six production and decay channels.  Our observables are each normalized to the corresponding SM expectation, so that in the SM their values are equal to 1.  We take the expected experimental uncertainties on these observables from the literature.  These expected uncertainties have been evaluated for 30~fb$^{-1}$ of integrated luminosity at one detector at the 14~TeV center-of-mass energy LHC, and include statistical uncertainties only.  We do not attempt to incorporate systematic uncertainties in the fit.  

Our first observable is the Higgs total width,
\begin{equation}
	\mathcal{O}_{1} = \frac{\Gamma_{\rm tot}}{\Gamma^{\rm SM}_{\rm tot}}
	= \frac{\bar g^2_W \Gamma^{\rm SM}_W + \bar g^2_Z \Gamma^{\rm SM}_Z 
	+ \Gamma_{\rm new}}{\Gamma^{\rm SM}_W + \Gamma^{\rm SM}_Z}.
	\label{eq:O1}
\end{equation}
Extraction of the Higgs total width from the $H \to ZZ \to 4\ell$ lineshape has been studied most thoroughly by CMS~\cite{Ball:2007zza}.  For an integrated luminosity of 30~fb$^{-1}$, CMS finds a statistical uncertainty on $\Gamma_{\rm tot}$ of 17.6\% for $M_H = 190$~GeV (see Sec.~10.2.1.7 of Ref.~\cite{Ball:2007zza}).  Because this uncertainty comes solely from the statistical precision in measuring the Gaussian width of the Higgs lineshape, it scales in the usual way as $(\int \! \mathcal{L})^{-1/2}$, where $\int \! \mathcal{L}$ is the integrated luminosity.  The uncertainty in the intrinsic $4\ell$ invariant mass resolution will enter as a systematic uncertainty, but has not been included in the analysis of Ref.~\cite{Ball:2007zza}.

Our next five observables are Higgs signal event rates studied for ATLAS in Ref.~\cite{Duhrssen}.  These comprise signal rates for Higgs production via gluon fusion and vector boson fusion, with decays to $WW$ and $ZZ$ final states (the vector boson fusion channel with decays to $WW$ is divided into two channels, one with $e \mu$ in the final state and the other with $ee$ or $\mu\mu$).  Because the signal and background selections in Ref.~\cite{Duhrssen} were done specifically for the purpose of Higgs coupling extraction, the author took care to account for the ``contamination'' of the signal in one production mode by Higgs production via the other production mode: for example, the event selections for Higgs production via gluon fusion contain some events in which the Higgs is produced by vector boson fusion, and vice versa.  Signal and background event numbers from Ref.~\cite{Duhrssen} are summarized in Table~\ref{tab:rates}.  Observables $\mathcal{O}_2$ and $\mathcal{O}_4$ select predominantly for Higgs production via gluon fusion, while $\mathcal{O}_3$, $\mathcal{O}_5$, and $\mathcal{O}_6$ select predominantly for Higgs production via vector boson fusion.
\begin{table}
\begin{tabular}{cccccccc}
\hline \hline
Observable & Process & $N_S$ & $N_B$ & Uncertainty & $R_{gg}$ & $R_{\rm VBF}$ \\
\hline
$\mathcal{O}_2$ & $H \to ZZ \to 4\ell$ & 68.1 ($gg$) + 11.2 (VBF) & 50.4 & 14.4\% & 85.9\% & 14.1\% \\
$\mathcal{O}_3$ & $H \to ZZ \to 4\ell$ & 15.2 (VBF) + 3.14 ($gg$) & 0.72 & 23.8\% & 17.1\% & 82.9\% \\
$\mathcal{O}_4$ & $H \to WW \to \ell\ell$ & 269 ($gg$) + 7.63 (VBF) & 428 & 9.60\% & 97.2\% & 2.8\% \\
$\mathcal{O}_5$ & $H \to WW \to e\mu$ & 78.0 (VBF) + 6.60 ($gg$) & 51.9 & 13.8\% & 7.8\% & 92.2\% \\
$\mathcal{O}_6$ & $H \to WW \to ee$, $\mu\mu$ & 73.2 (VBF) + 5.70 ($gg$) & 55.8 & 14.7\% & 7.2\% & 92.8\% \\
\hline \hline
\end{tabular}
\caption{Higgs signal and background rates for observables $\mathcal{O}_2$ through $\mathcal{O}_6$, from Ref.~\cite{Duhrssen}.  Here $\ell$ includes $e$ and $\mu$.  Event numbers are for 30~fb$^{-1}$.  The statistical uncertainty on the signal rate is given by $\sqrt{N_S + N_B}/N_S$.  We also quote the fractions of the signal events coming from gluon fusion ($R_{gg}$) and vector boson fusion ($R_{\rm VBF}$) for each rate observable.}
\label{tab:rates}
\end{table}
For our fit, we define these observables as the rate for the selected process normalized to the SM rate:
\begin{equation}
	\mathcal{O}_i = \frac{{\rm Rate}_i}{{\rm Rate}^{\rm SM}_i}
	= R_{gg}^i \frac{\sigma(gg \to H) {\rm BR}_i}{\sigma^{\rm SM}(gg \to H) {\rm BR}^{\rm SM}_i}
	+ R_{\rm VBF}^i \frac{\sigma({\rm VBF} \to H) {\rm BR}_i}{\sigma^{\rm SM}({\rm VBF} \to H) {\rm BR}^{\rm SM}_i}, \qquad i = 2 \ldots 6,
\end{equation}
where $R_{gg}^i$ and $R_{\rm VBF}^i$ are the fractions of signal events coming from gluon fusion and vector boson fusion, respectively, as given in Table~\ref{tab:rates}.  Using Eq.~(\ref{eq:params}), these observables can be expressed in terms of our fit parameters as
\begin{eqnarray}
	\mathcal{O}_i &=& \left\{ R_{gg}^i \bar g_g^2 + R_{\rm VBF}^i \left[ 0.73 \, \bar g_W^2 + (1 - 0.73) \bar g_Z^2 \right] \right\} \bar g_Z^2 \frac{\Gamma_{\rm tot}^{\rm SM}}{\Gamma_{\rm tot}}, \qquad 
	i = 2, 3, \nonumber \\
	\mathcal{O}_i &=& \left\{ R_{gg}^i \bar g_g^2 + R_{\rm VBF}^i \left[ 0.73 \, \bar g_W^2 + (1 - 0.73) \bar g_Z^2 \right] \right\} \bar g_W^2 \frac{\Gamma_{\rm tot}^{\rm SM}}{\Gamma_{\rm tot}}, \qquad 
	i = 4, 5, 6,
\end{eqnarray}
where the ratio $\Gamma_{\rm tot}/\Gamma_{\rm tot}^{\rm SM}$ has been given in terms of the fit parameters in Eq.~(\ref{eq:O1}).

Our last observable comes from a CMS analysis of Higgs production in vector boson fusion with decays to $WW$ in the channel $jj\ell \nu$~\cite{CMSnote}.  The analysis in Ref.~\cite{CMSnote} did not take into account any ``contamination'' of the signal by Higgs production via gluon fusion; we take this at face value for the purpose of our fit.  The observable is the ratio of the signal rate to its SM value, which can be expressed as
\begin{equation}
	\mathcal{O}_7 = \left[ 0.73 \, \bar g_W^2 + (1 - 0.73) \bar g_Z^2 \right] \bar g_W^2 \frac{\Gamma_{\rm tot}^{\rm SM}}{\Gamma_{\rm tot}},
\end{equation}
where again the ratio $\Gamma_{\rm tot}/\Gamma_{\rm tot}^{\rm SM}$ has been given in terms of the fit parameters in Eq.~(\ref{eq:O1}).  After cuts, Ref.~\cite{CMSnote} found a signal cross section of 2.340~fb and a background cross section of 1.567~fb, leading to a statistical uncertainty of 15.4\% in 30~fb$^{-1}$.

We note that the statistical power of the $H \to ZZ$ channels could be improved by including $\ell \ell \nu \nu$ and $\ell \ell bb$ final states.  These have been studied as LHC Higgs discovery channels; ATLAS~\cite{ATLHproj} has studied both channels for $M_H \geq 200$~GeV, while CMS~\cite{CMSHproj} has studied $\ell\ell\nu\nu$ for $M_H \geq 200$~GeV and $\ell\ell bb$ for $M_H \geq 300$~GeV.

In order to evaluate the precision with which these LHC measurements would be able to constrain the Higgs couplings, we compute a $\chi^2$ for each choice of the parameters $(\bar g_W^2, \bar g_Z^2, \bar g_g^2, \Gamma_{\rm new})$ according to
\begin{equation}
	\chi^2 = \sum_{i=1}^7 \frac{\left(\mathcal{O}_i - \mathcal{O}_i^{\rm SM}\right)^2}{\sigma_i^2},
\end{equation}
where our observables are normalized such that $\mathcal{O}_i^{\rm SM} = 1$ and $\sigma_i$ is the (fractional) uncertainty on observable $\mathcal{O}_i$.  This amounts to assuming that the measured values of all the observables are equal to the SM prediction; thus $\chi^2 = 0$ at the SM point $(\bar g_W^2, \bar g_Z^2, \bar g_g^2, \Gamma_{\rm new}) = (1, 1, 1, 0)$.  Note that we ignore correlated uncertainties, the most important of which we expect to be the luminosity uncertainty (which affects $\mathcal{O}_2$ through $\mathcal{O}_7$), the theoretical uncertainty on the gluon fusion cross section (which affects $\mathcal{O}_2$ through $\mathcal{O}_6$), and the theoretical uncertainty on the weak boson fusion cross section (which affects $\mathcal{O}_2$ through $\mathcal{O}_7$).  Uncertainty on the background normalization for the channels with $H \to WW$ may also be important, and affects $\mathcal{O}_4$ through $\mathcal{O}_7$.

We perform two sets of fits: one with three free parameters $(\bar g_V^2, \bar g_g^2, \Gamma_{\rm new}/\Gamma_{\rm tot}^{\rm SM})$, in which we take $\bar g_W^2 = \bar g_Z^2 \equiv \bar g_V^2$; and one with four free parameters $(\bar g_W^2, R, \bar g_g^2, \Gamma_{\rm new}/\Gamma_{\rm tot}^{\rm SM})$, where 
\begin{equation}
	R \equiv \frac{\bar g_Z^2}{\bar g_W^2}.
\end{equation}
We choose $R$ as the fourth variable instead of $\bar g_Z^2$ because we are most interested in how well the LHC will be able to test the prediction $\bar g_W = \bar g_Z$ of extended Higgs models that contain only Higgs doublets and singlets.  When presenting our results we normalize the new contribution to the Higgs total width to the SM Higgs total width, thus plotting limits on the dimensionless quantity $\Gamma_{\rm new}/\Gamma_{\rm tot}^{\rm SM}$.

In each case we scan over the free parameters and compute $\chi^2$ at each parameter point.  We then project the $\chi^2$ distribution down onto two or one parameters by marginalizing over the undisplayed parameters---in other words, we find the smallest value of $\chi^2$ that can be obtained by varying the undisplayed parameters.  In each case we plot the resulting 1$\sigma$, 2$\sigma$, and 3$\sigma$ constraints.  Projecting onto one parameter, this corresponds to $\chi^2 = 1$, 4, and 9.  Projecting onto two parameters, this corresponds to $\chi^2 = 2.296$, 6.180, and 11.829. 

We will also display results extrapolated to an integrated luminosity of 100~fb$^{-1}$.  Because all our uncertainties are statistical, we can do this simply by scaling all uncertainties by $(\int \! \mathcal{L})^{-1/2}$, i.e., by multiplying the uncertainties for 30~fb$^{-1}$ by $\sqrt{3/10}$.  All results presented assume data from only one detector---combining data from ATLAS and CMS would effectively double the statistics.

\section{Results}
\label{sec:results}

\begin{figure}
\resizebox{\textwidth}{!}{\includegraphics{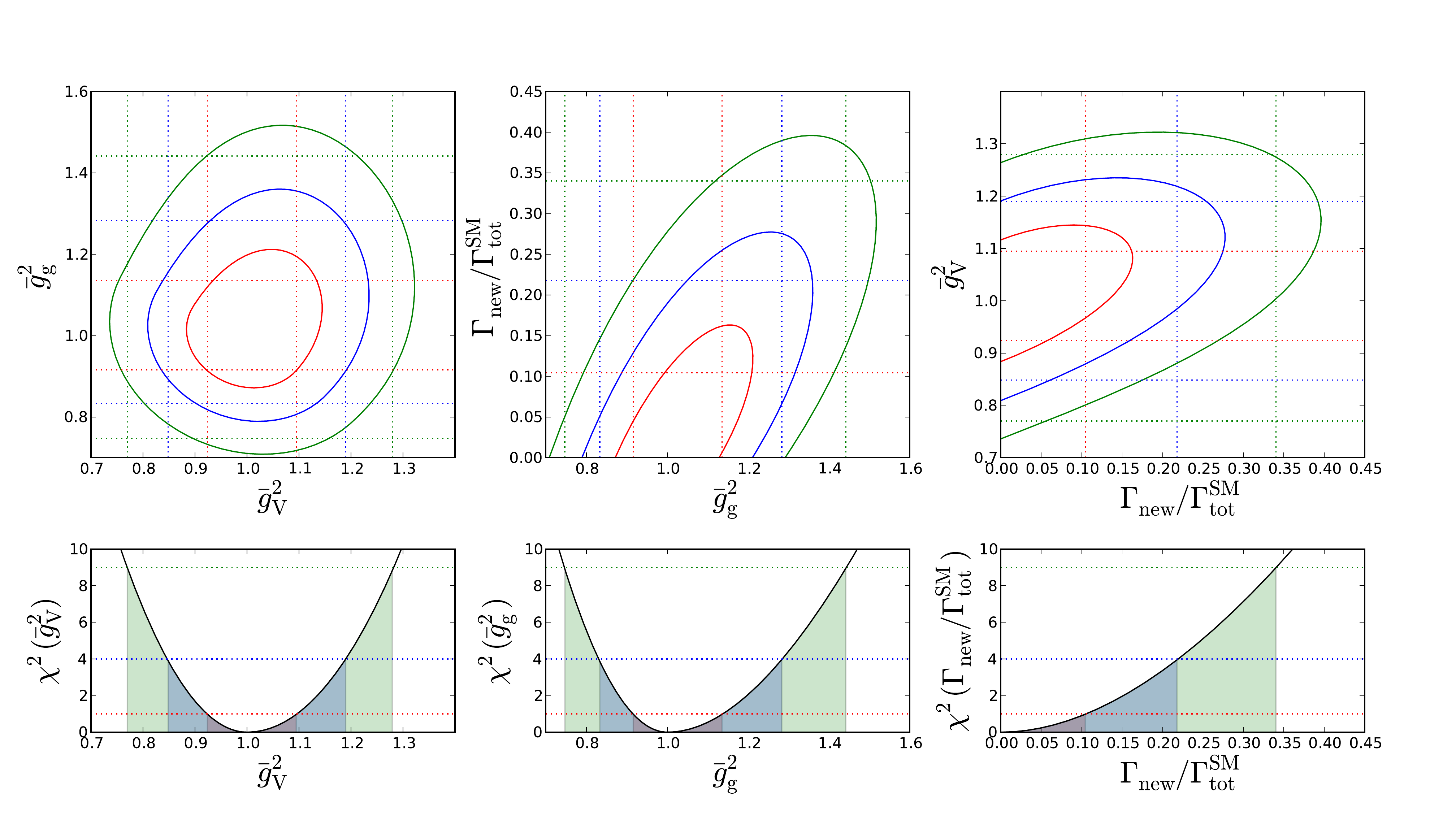}}
\caption{Results of the three-parameter scan over $(\bar g_V^2, \bar g_g^2, \Gamma_{\rm new}/\Gamma_{\rm tot}^{\rm SM})$, with $\bar g_W^2 = \bar g_Z^2 \equiv \bar g_V^2$, for $M_H = 190$ GeV and 30 fb$^{-1}$ of integrated luminosity.  The upper plots show the $1\sigma$, $2\sigma$, and $3\sigma$ regions for each pair of parameters with the $\chi^2$ marginalized over the remaining parameter.  The lower plots show the marginalized $\chi^2$ distributions for each parameter, with the shaded regions indicating the $1\sigma$, $2\sigma$, and $3\sigma$ ranges.  These ranges for each parameter are indicated on the upper plots by the straight dotted lines.}
\label{fig:3param-30ifb}
\end{figure}

We begin by scanning over the three-dimensional parameter space of $(\bar g_V^2, \bar g_g^2, \Gamma_{\rm new}/\Gamma_{\rm tot}^{\rm SM})$, with $\bar g_V^2 \equiv \bar g_W^2 = \bar g_Z^2$, using the uncertainties corresponding to 30~fb$^{-1}$ of integrated luminosity to calculate the $\chi^2$.  Results are shown in Fig.~\ref{fig:3param-30ifb}, in which we plot the projections of the $\chi^2$ distribution onto the three pairs of variables as well as onto each individual variable.  
The solid contours correspond to the 1$\sigma$, 2$\sigma$, and 3$\sigma$ constraints.  The straight dashed lines show the 1$\sigma$, 2$\sigma$, and 3$\sigma$ ranges of the individual parameters.  Because we compute the $\chi^2$ assuming that the ``observed'' values of the inputs are equal to their SM predictions, the minimum value of the $\chi^2$ is zero and occurs at $(\bar g_V^2, \bar g_g^2, \Gamma_{\rm new}/\Gamma_{\rm tot}^{\rm SM}) = (1, 1, 0)$.  We see that $\bar g_V^2$ and $\bar g_g^2$ can be measured with about 8.5\% and 11\% precision respectively, and $\Gamma_{\rm new}/\Gamma_{\rm tot}^{\rm SM}$ is bounded to be below about 22\% at the 2$\sigma$ level.  1$\sigma$, 2$\sigma$, and 3$\sigma$ ranges for each of the parameters are summarized in Table~\ref{tab:3param-ranges}.

\begin{table}
\begin{tabular}{ccccc}
\hline\hline
Parameter & $\int \! \mathcal{L}$ & 1$\sigma$ & 2$\sigma$ & 3$\sigma$ \\
\hline
$\bar g_V^2$ & 30 fb$^{-1}$ & 0.924--1.095 & 0.848--1.190 & 0.770--1.280 \\
& 100 fb$^{-1}$ & 0.960--1.053 & 0.918--1.103 & 0.876--1.154 \\
$\bar g_g^2$ & 30 fb$^{-1}$ & 0.916--1.136 & 0.833--1.283 & 0.747--1.442 \\
& 100 fb$^{-1}$ & 0.956--1.071 & 0.909--1.150 & 0.862--1.229 \\
$\Gamma_{\rm new}/\Gamma_{\rm tot}^{\rm SM}$ & 30 fb$^{-1}$ & 0--0.104 & 0--0.218 & 0--0.340 \\
& 100 fb$^{-1}$ & 0--0.056 & 0--0.115 & 0--0.176 \\
\hline\hline
\end{tabular}
\caption{Parameter ranges allowed at 1$\sigma$, 2$\sigma$, and 3$\sigma$ by the three-parameter fit, with 30~fb$^{-1}$ or 100~fb$^{-1}$ of integrated luminosity.  Here $\bar g_W^2 = \bar g_Z^2 \equiv \bar g_V^2$.}
\label{tab:3param-ranges}
\end{table}

\begin{figure}
\resizebox{\textwidth}{!}{\includegraphics{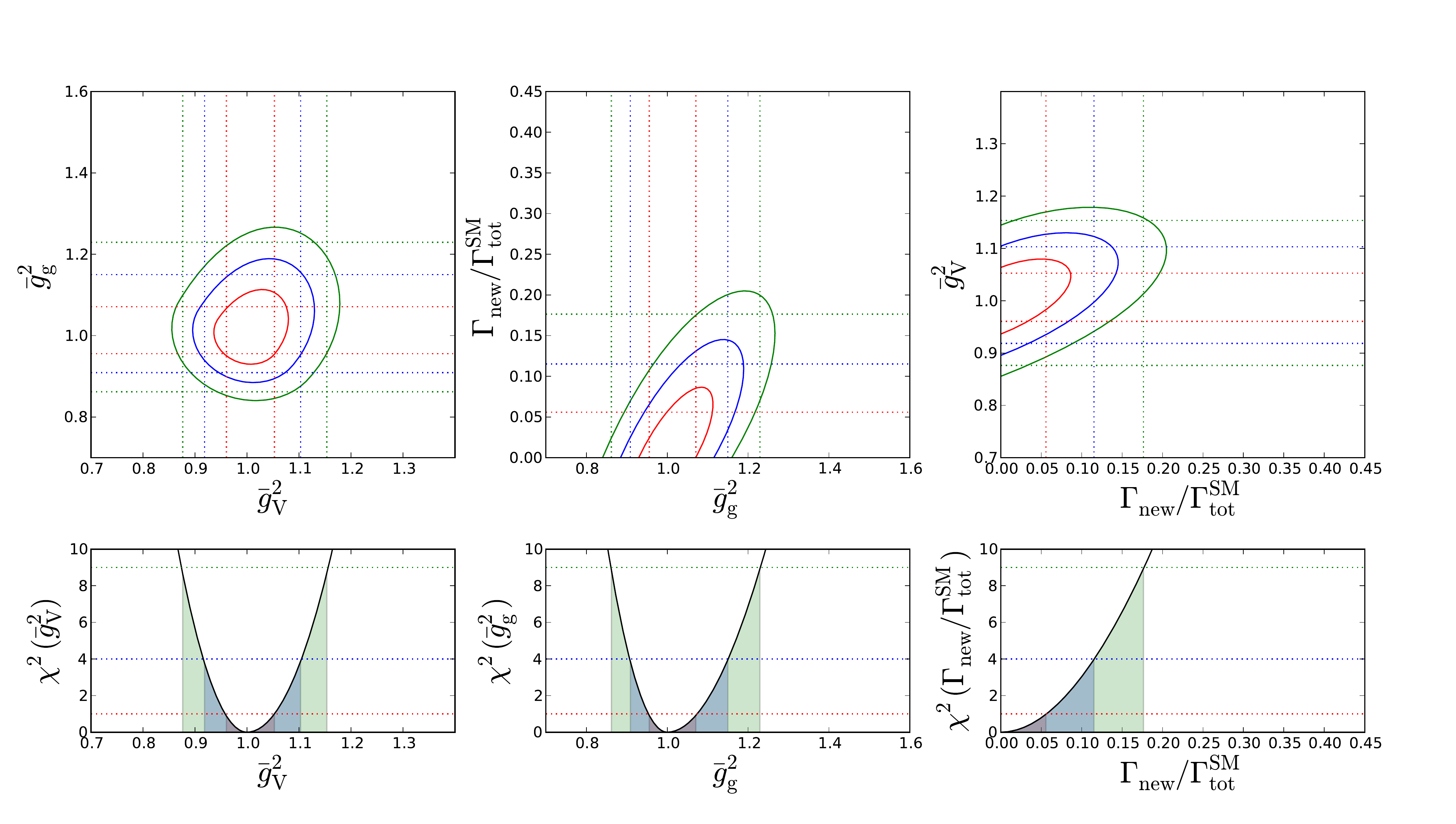}}
\caption{As in Fig.~\ref{fig:3param-30ifb} but for 100~fb$^{-1}$ of integrated luminosity.}
\label{fig:3param-100ifb}
\end{figure}

Scaling the statistical uncertainties to an integrated luminosity of 100~fb$^{-1}$ tightens the constraints as shown in Fig.~\ref{fig:3param-100ifb}.  This reduces the uncertainties on $\bar g_V^2$ and $\bar g_g^2$ to about 4.6\% and 5.8\% respectively, and lowers the 2$\sigma$ upper limit on $\Gamma_{\rm new}/\Gamma_{\rm tot}^{\rm SM}$ to about 12\%.  We emphasize here that at this level of statistical precision, we expect that systematic uncertainties, which have not been included in our fit, will begin to play a significant role.

\begin{figure}
\resizebox{\textwidth}{!}{\includegraphics{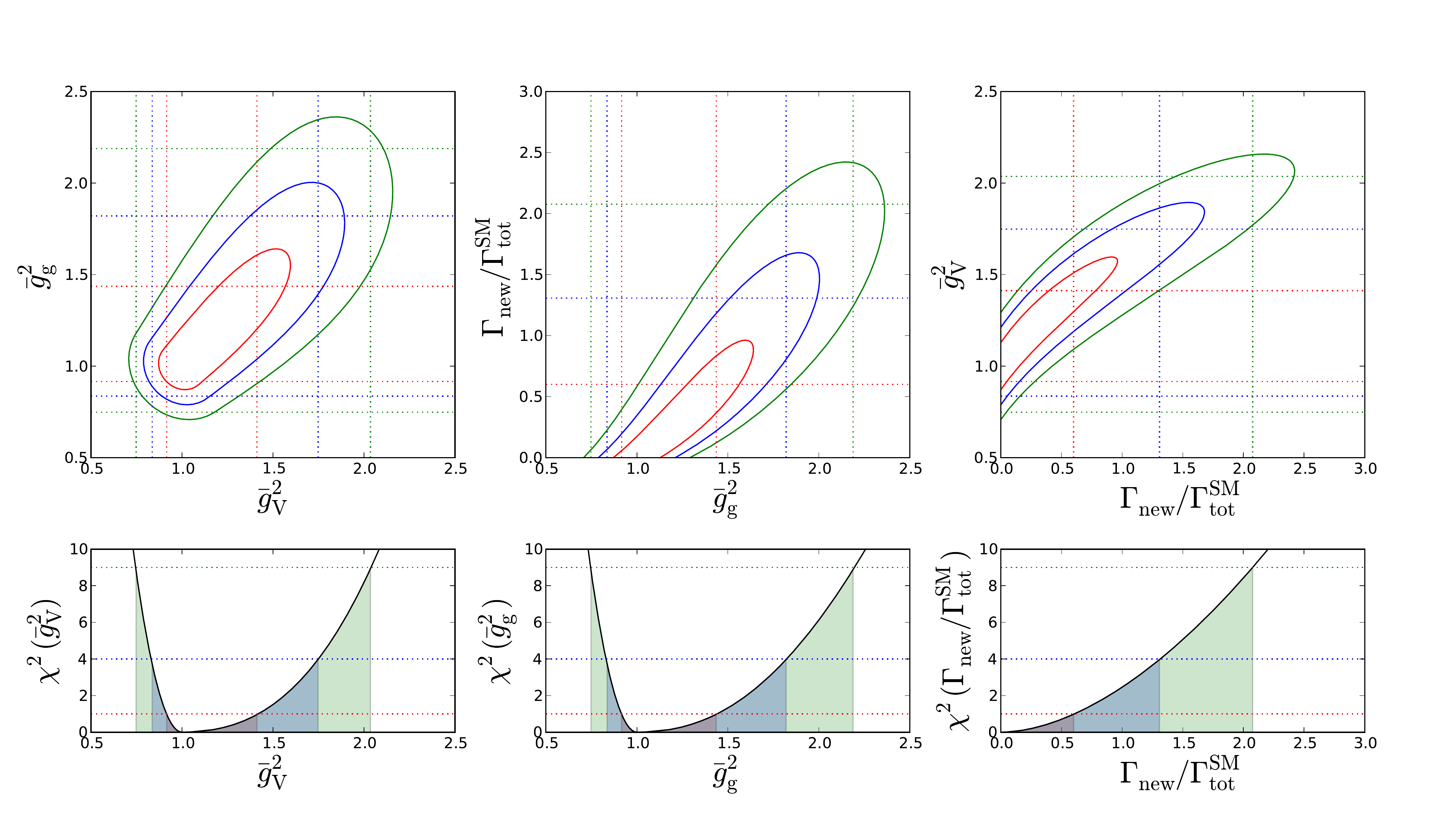}}
\caption{As in Fig.~\ref{fig:3param-30ifb} but with the uncertainty on the measurement of $\Gamma_{\rm tot}/\Gamma_{\rm tot}^{\rm SM}$ artificially set to 100\%.  Note the expanded range of the axes compared to Fig.~\ref{fig:3param-30ifb}.}
\label{fig:badwidth}
\end{figure}

Notice in both Figs.~\ref{fig:3param-30ifb} and~\ref{fig:3param-100ifb} that $\Gamma_{\rm new}/\Gamma_{\rm tot}^{\rm SM}$ is positively correlated with both $\bar g_g^2$ and $\bar g_V^2$.  This is a manifestation of the flat direction discussed in the introduction, which has been lifted by the inclusion of the direct measurement of the Higgs total width in the fit.  To illustrate the importance of the total width measurement, we redo the fit for 30~fb$^{-1}$ with the uncertainty on the total width measurement artificially inflated from 17.6\% to 100\%.  Results are shown in Fig.~\ref{fig:badwidth}.

\begin{figure}
\resizebox{\textwidth}{!}{\includegraphics{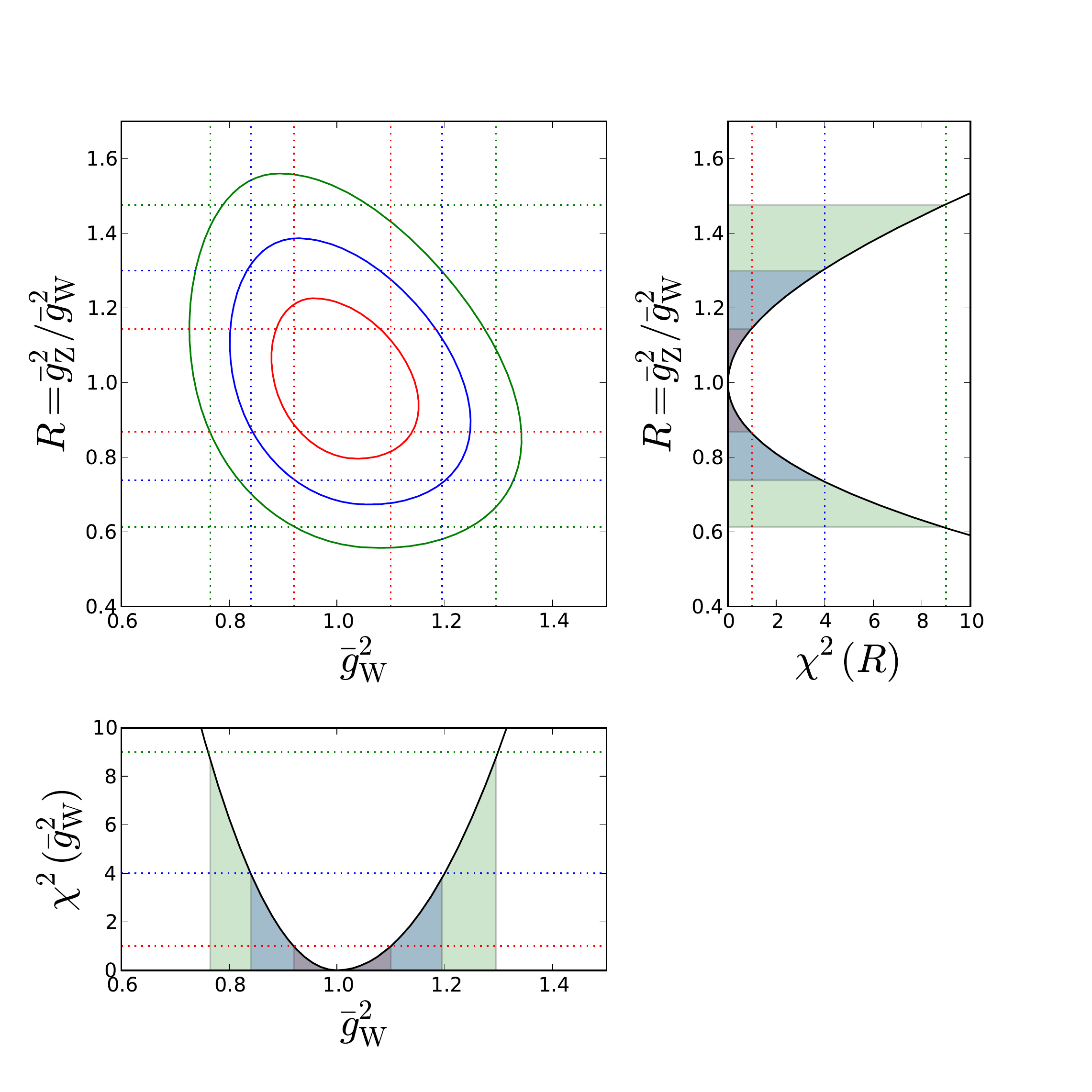}
\hspace*{-1cm}
\includegraphics{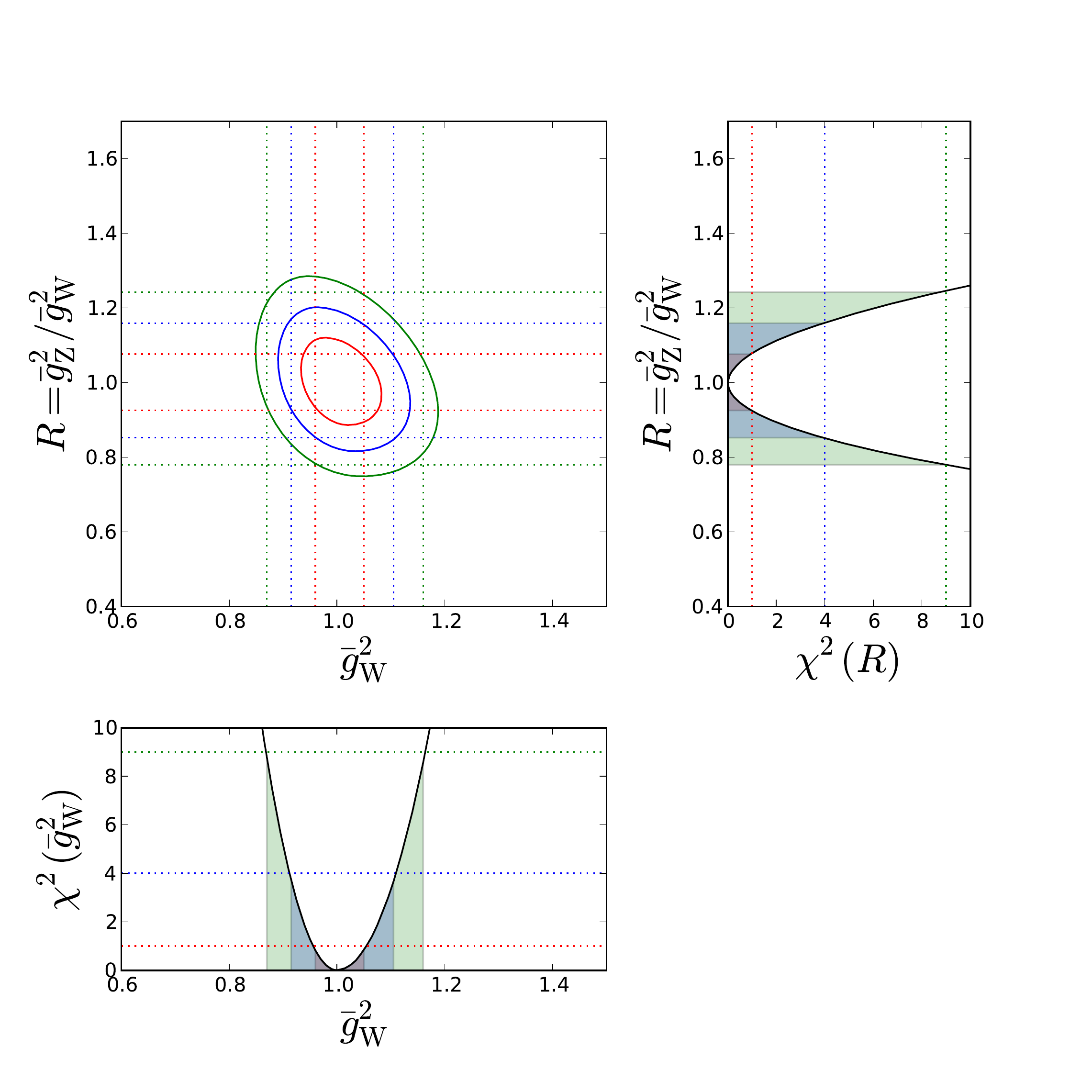}}
\caption{Results of the four-parameter scan over $(\bar g_W^2, R, \bar g_g^2, \Gamma_{\rm new}/\Gamma_{\rm tot}^{\rm SM})$, for $M_H = 190$~GeV and integrated luminosities of 30~fb$^{-1}$ (left) and 100~fb$^{-1}$ (right).  We show the $1\sigma$, $2\sigma$, and $3\sigma$ regions for the pair of parameters $(\bar g_W^2, R)$ with the $\chi^2$ marginalized over the remaining two parameters.  We also show the marginalized $\chi^2$ distributions for each of these two parameters, with the shaded regions indicating the $1\sigma$, $2\sigma$, and $3\sigma$ ranges.  These ranges for $\bar g_W^2$ and $R$ are indicated on the two-dimensional plots by the straight dotted lines.}
\label{fig:4param}
\end{figure}

Finally, we scan over the four-dimensional parameter space of $(\bar g_W^2, R, \bar g_g^2, \Gamma_{\rm new}/\Gamma_{\rm tot}^{\rm SM})$, with $R \equiv \bar g_Z^2/ \bar g_W^2$.  The resulting constrains on $\bar g_W^2$ and $R$ are shown for 30~fb$^{-1}$ and 100~fb$^{-1}$ in the left and right panels of Fig.~\ref{fig:4param}.  
We see that $R$ can be measured with a precision of about 14\% with 30~fb$^{-1}$, improving to about 8\% with 100~fb$^{-1}$ (again we emphasize that these uncertainties are statistical only).  The precisions on the other three parameters are comparable to those in the three-parameter fit.  1$\sigma$, 2$\sigma$, and 3$\sigma$ ranges for each of the parameters are summarized in Table~\ref{tab:4param-ranges}.  Note that due to the finite spacing of our four-parameter scan grids, the quoted uncertainties are only accurate to within about 2 percentage points.

\begin{table}
\begin{tabular}{ccccc}
\hline\hline
Parameter & $\int \! \mathcal{L}$ & 1$\sigma$ & 2$\sigma$ & 3$\sigma$ \\
\hline
$\bar g_W^2$ & 30 fb$^{-1}$ & 0.92--1.10 & 0.84--1.20 & 0.77--1.30 \\
& 100 fb$^{-1}$ & 0.96--1.05 & 0.92--1.11 & 0.87--1.16 \\
$R \equiv \bar g_Z^2/\bar g_W^2$ & 30 fb$^{-1}$ & 0.87--1.14 & 0.74--1.30 & 0.61--1.48 \\
& 100 fb$^{-1}$ & 0.93--1.08 & 0.85--1.16 & 0.78--1.24 \\
$\bar g_g^2$ & 30 fb$^{-1}$ & 0.92--1.13 & 0.84--1.28 & 0.75--1.44 \\
& 100 fb$^{-1}$ & 0.96--1.07 & 0.91--1.15 & 0.86--1.23 \\
$\Gamma_{\rm new}/\Gamma_{\rm tot}^{\rm SM}$ & 30 fb$^{-1}$ & 0--0.10 & 0--0.21 & 0--0.34 \\
& 100 fb$^{-1}$ & 0--0.05 & 0--0.11 & 0--0.17 \\
\hline\hline
\end{tabular}
\caption{Parameter ranges allowed at 1$\sigma$, 2$\sigma$, and 3$\sigma$ by the four-parameter fit, with 30~fb$^{-1}$ or 100~fb$^{-1}$ of integrated luminosity.  Note that due to the finite spacing of our four-parameter scan grids, the quoted parameter range boundaries are only accurate to within about 0.02.}
\label{tab:4param-ranges}
\end{table}

For comparison, we can estimate the precision with which $R$ can be measured by taking the ratio of Higgs signal rates with a common production mode but with decays to $ZZ$ versus $WW$.  In such a ratio the production cross section and Higgs total width cancel, leaving the ratio of the partial widths of $H \to ZZ$ and $H \to WW$, each of which is proportional to the square of the respective coupling.  In particular, for Higgs production via gluon fusion we can take the ratio of $\mathcal{O}_2$ and $\mathcal{O}_4$, while for Higgs production via vector boson fusion we can take the ratio of $\mathcal{O}_3$ and a combination of $\mathcal{O}_5$, $\mathcal{O}_6$, and $\mathcal{O}_7$.  Propagating the uncertainties in the usual way, we find an uncertainty on $R$ of about 14\% with 30~fb$^{-1}$ and 7.8\% with 100~fb$^{-1}$, very consistent with the results of the $\chi^2$ scan.  This ratio method is inexact due to the fact that the channels with different final states have different amounts of ``contamination'' of the desired production mode by the other production mode; i.e., $R^i_{gg}$ and $R^i_{\rm VBF}$ are different for the $\mathcal{O}_i$ in the numerator and the denominator, so that the parameter dependence in the production cross sections does not cancel perfectly.

\section{Discussion and conclusions}
\label{sec:conclusions}

In this paper we showed that truly model-independent Higgs coupling measurements can be made at the LHC when the Higgs total width is experimentally accessible.  For a SM-like Higgs, this occurs for $M_H \gtrsim 190$~GeV.  At such high masses, however, the only accessible Higgs decay modes in the SM are to $WW$ and $ZZ$, with production via gluon fusion or vector boson fusion.  Thus the LHC measurements provide access to the Higgs couplings to $WW$ and $ZZ$, the effective coupling to gluons, and any new contribution to the Higgs total width beyond the decay widths to $WW$ and $ZZ$.  

Using existing LHC studies for a Higgs mass of 190~GeV at 14~TeV center-of-mass energy and considering only statistical uncertainties, we found that with 30~fb$^{-1}$ at one detector the Higgs coupling-squared to vector bosons (assuming the ratio of $WW$ and $ZZ$ couplings is equal to its SM value) can be measured with an uncertainty of about 8.5\%, the Higgs effective coupling-squared to gluons can be measured with an uncertainty of about 11\%, and a new, non-SM component of the Higgs total width can be constrained at the 2$\sigma$ level to be below about 22\% of the SM Higgs total width at this mass.  With 100~fb$^{-1}$ at one detector the coupling-squared uncertainties decrease to about 4.6\% and 5.8\%, respectively, and the 2$\sigma$ upper limit on a new contribution to the Higgs total width decreases to about 12\% of the SM Higgs total width.  At this level of precision we expect that systematic uncertainties will be important.

To give a sense of the usefulness of our constraint on a new contribution to the Higgs total width, consider Higgs decays to invisible final states such as dark matter particles.  ATLAS has studied the sensitivity to invisible Higgs decays in vector boson fusion with 30~fb$^{-1}$ at 14~TeV.  For $M_H = 190$~GeV they find a 95\% confidence level sensitivity to $\xi^2 \equiv [\sigma({\rm VBF} \to H)/\sigma^{\rm SM}({\rm VBF} \to H)] \times {\rm BR}(H \to {\rm invis.})$ of about 15\% not including systematic uncertainties, which rises to 60\% including systematics~\cite{Aad:2009wy}.
Their dominant systematic uncertainty comes from the 10\% background normalization uncertainty due to the theoretical uncertainty in the shape of the angular distribution of the two jets in the dominant $Wjj$ and $Zjj$ backgrounds; using a next-to-leading order calculation for $Wjj$ and $Zjj$ should reduce this by a factor of two, improving the 95\% confidence level sensitivity to $\xi^2$ to roughly 40\%.  For SM-like Higgs couplings to $WW$ and $ZZ$, this more optimistic limit corresponds to a new invisible component of the Higgs total width of $\Gamma_{\rm new}/\Gamma_{\rm tot}^{\rm SM} \simeq 65\%$, which should be easily detectable (but not identifiable as an invisible decay) using our methods.

How do our results compare with those achievable at the ILC?  Results of multiple ILC Higgs coupling studies at a variety of Higgs masses are summarized in Ref.~\cite{Battaglia:2002av}, which incorporates results for $e^+e^-$ center-of-mass energies of 350--800~GeV and integrated luminosities of 500--1000~fb$^{-1}$.  While $M_H = 190$~GeV was not explicitly studied in Ref.~\cite{Battaglia:2002av}, we can obtain coupling uncertainties for this mass from Fig.~5 of Ref.~\cite{Battaglia:2002av}, which interpolates between higher and lower studied mass points.  From this we read off an ILC uncertainty on $\bar g_W^2$ of about 3\%, to be compared with our LHC statistical precision of 8.5\% (4.6\%) with 30~fb$^{-1}$ (100~fb$^{-1}$).  ILC has no access to $\bar g_g^2$ for this Higgs mass due to the extreme suppression of the $H \to gg$ branching fraction.  The direct ILC measurement of the Higgs total width from the $H \to ZZ \to 4\ell$ lineshape is limited by statistics.  However, it was shown in Ref.~\cite{Richard:2007ru} that prospects seem good to measure the Higgs total width to better than 10\% precision from the Higgs recoil mass lineshape in $e^+e^- \to ZH$, running at a center-of-mass energy not too far above the threshold for $ZH$ associated production.

Of greater importance, however, are the complementary ILC measurements of the Higgs couplings to $b \bar b$ and $t \bar t$ that cannot be measured at the LHC for the relatively high Higgs mass considered here.  Despite the very small branching fraction of $H \to b \bar b$ at $M_H = 190$~GeV, $\bar g_b^2$ can be measured to about 14\% precision at the ILC~\cite{Battaglia:2002av}, where $\bar g_b$ is a multiplicative factor parameterizing the shift in the $H b \bar b$ coupling from its SM value.  A detailed study of the measurement of the Higgs coupling to $t \bar t$ from the process $e^+e^- \to t \bar t H$, $H \to WW$ was made in Ref.~\cite{Gay:2006vs}, assuming 1000~fb$^{-1}$ at an $e^+e^-$ center-of-mass energy of 800~GeV.  Interpolating the results of Ref.~\cite{Gay:2006vs} to $M_H = 190$~GeV, we can read off an uncertainty on $\bar g_t^2$ (defined analogously to $\bar g_b^2$) of about 21\% (24\%) for a residual background normalization uncertainty of 5\% (10\%).  Combining this measurement of $\bar g_t^2$ with the LHC measurement of $\bar g_g^2$ allows one to probe contributions to the effective $Hgg$ coupling due to particles other than the top quark.  

We note that all the ILC coupling measurement precisions quoted here have been extracted with various model assumptions applied, rather than from a fully model-independent fit to ILC observables.

We now discuss systematic uncertainties.  We expect the most important systematic uncertainties to be the luminosity uncertainty, theoretical and parton distribution function (PDF) uncertainties on the gluon fusion and vector boson fusion cross sections, the background normalization uncertainty particularly in the $H \to WW$ channels, and the uncertainty in the experimental resolution of the $4\ell$ invariant mass used in the Higgs width determination.  We comment on these as follows:
\begin{itemize}
\item Previous studies of Higgs coupling extraction from LHC data have assumed a luminosity uncertainty of 5\%~\cite{Zeppenfeld:2000td,Duhrssen,Duhrssen:2004cv,Lafaye:2009vr}.  Recent ATLAS~\cite{ATLASlumi} and CMS~\cite{CMSlumi} luminosity determinations from data collected in 2010 have achieved uncertainties of 3.4\% and 4.0\%, respectively.  Normalizing LHC rate measurements to the inclusive $W$ or $Z$ production rates, which are now known to next-to-next-to-leading order in QCD and with next-to-leading order electroweak corrections leading to a theoretical uncertainty below 1\%, could replace the luminosity uncertainty with the small statistical and PDF uncertainties on the $W$ or $Z$ rate measurement.

\item Recent improvements in the calculation of the gluon fusion Higgs production cross section have significantly reduced the theoretical uncertainty.  The LHC Higgs Cross Section Working Group~\cite{Dittmaier:2011ti} quotes a conservative theoretical uncertainty of 8.4\% and a PDF uncertainty of 6.8\% for $M_H = 190$~GeV at the 14~TeV LHC (these numbers become 8.3\% and 7.9\% at 7~TeV center-of-mass energy).  Other groups are more aggressive on the theory uncertainty; for example, Ref.~\cite{Ahrens:2010rs} combines the most up-to-date calculations including renormalization-group improvement of the QCD corrections at next-to-next-to-next-to-leading logarithmic accuracy as well as next-to-leading order electroweak corrections, and finds a remaining scale uncertainty of only 1.3\% (1.5\%) for 14~TeV (7~TeV) center-of-mass energy.  The uncertainty on the gluon PDF is also expected to improve as LHC data become available to be incorporated into the global PDF fits.

\item The cross section for Higgs production via vector boson fusion is now known through next-to-next-to-leading order in QCD and includes next-to-leading order electroweak corrections.  The remaining theoretical scale uncertainty at the LHC is a mere 0.3\% (0.2\%) and the PDF uncertainty is 2.5\% (2.6\%) at 14~TeV (7~TeV)~\cite{Dittmaier:2011ti}.

\item For data-driven background determination, the background normalization uncertainty comes from the statistical uncertainty in the background control samples as well as the theoretical uncertainty in the extrapolation from the control regions into the signal region.  In previous studies the most important of these has been the $WW$ background, for which the number of events in the control region is no larger than in the signal region and the extrapolation error from shape uncertainty has been taken as 5\%~\cite{Duhrssen:2004cv}.

\item The two-lepton invariant mass resolution should be well-calibrated from the $Z$ peak.  We expect that the four-lepton invariant mass resolution can be determined from this.  We are not aware of a discussion of this uncertainty in the literature.
\end{itemize}

The Higgs lineshape and production rate measurement studies that we used in this analysis were all done for an LHC center-of-mass design energy of 14~TeV.  The LHC is currently running at a center-of-mass energy of 7~TeV, however, and this lower-energy running is anticipated to continue to the end of 2012.  The LHC experiments have already collected more than 1~fb$^{-1}$ of data, and may be able to collect several~fb$^{-1}$ at this lower center-of-mass energy by the end of 2012.  Because of this, it is interesting to consider how much data would be needed at 7~TeV to achieve the uncertainties obtained here for 14~TeV.  For a mass of 190~GeV, Higgs signal cross sections in both gluon fusion and vector boson fusion are smaller by a factor of 3.8~\cite{Dittmaier:2011ti} at 7~TeV compared to their values at 14~TeV.  Combining data from both ATLAS and CMS, each experiment would need almost 60~fb$^{-1}$ at 7~TeV to together collect the same number of signal events as used in our analysis of 30~fb$^{-1}$ at one detector.  A more quantitative estimate would require information on the background cross sections at 7~TeV, and possibly involve a new optimization of the signal selections.

\begin{acknowledgments}
We thank Alain Bellerive for helpful discussions of statistics.
H.E.L.\ was supported in part by the Natural Sciences and Engineering
Research Council of Canada.
\end{acknowledgments}


\end{document}